# Evidence of zero point fluctuation of vortices in a very weakly pinned a-MoGe thin film

Surajit Dutta[a], Indranil Roy[a], John Jesudasan[a], Subir Sachdev[b] and Pratap Raychaudhuri[a*]

[a]*Tata Institute of Fundamental Research, Homi Bhabha Road, Colaba, Mumbai 400005, India.*

[b]*Department of Physics, Harvard University, Cambridge MA 02138, USA.*

In a Type II superconductor, the vortex core behaves like a normal metal. Consequently, the single-particle density of states in the vortex core of a conventional Type II superconductor remains either flat or (for very clean single crystals) exhibits a peak at zero bias due to the formation of Caroli-de Gennes-Matricon bound state inside the core. Here we report an unusual observation from scanning tunneling spectroscopy measurements in a weakly pinned thin film of the conventional s-wave superconductor *a*-MoGe, namely, that a soft gap in the local density of states continues to exist even at the center of the vortex core. We ascribe this observation to rapid fluctuation of vortices about their mean position that blurs the boundary between the gapless normal core and the gapped superconducting region outside. Analyzing the data as a function of magnetic field we show that the variation of fluctuation amplitude as a function of magnetic field is consistent with quantum zero-point motion of vortices.

[*] pratap@tifr.res.in

## I. Introduction

In Type II superconductors, magnetic field penetrates inside the sample above the lower critical field ($H_{c1}$) in the form of quantized flux tubes called vortices[1], each carrying a magnetic flux of $\Phi_0 = \frac{h}{2e} = 2.07 \times 10^{-15}\ Wb$. In the diamagnetic background, the magnetic flux in a vortex is sustained by a circulating supercurrent density ($J_s$) that falls off inversely with distance from the center of the vortex, $r$, as $\boldsymbol{J_s} \propto 1/r$. The divergence in $\boldsymbol{J_s}$ as $r \rightarrow 0$ is cut off through the formation of a normal metal core at the center of the vortex of the size of the superconducting coherence length, $\xi$. In the presence of an external transport current, vortices are subjected to a Lorentz force which can cause them to move and cause dissipation in the system[2]. Since the dissipation caused by a moving vortex is contributed by its normal metal core, it is of practical interest to have a detailed understanding of the electronic structure inside the vortex core (VC). Due to the confinement of normal electrons in the VC, Caroli-de Gennes-Matricon (CdGM) predicted that they will form bound states with energy with respect to the Fermi energy, $E_F$, at $\varepsilon_n = \pm \frac{n\Delta^2}{E_F}$ ( $n$ = 1/2, 3/2, 5/2, …), where Δ is the superconducting energy gap in zero field[3]. However, due to the smallness of $\frac{\Delta}{E_F}$, individual CdGM states are difficult to observe. So far, individual CdGM peaks have only been observed in some iron based superconductors where $E_F$ is small[4,5]. On the other hand, in conventional superconductors, the accumulated density of state from many CdGM states manifest in the form of a peak in the local density of states (LDOS) centered about $E_F$, that smoothly decays when one moves away from the center of the vortex core[6]. However, the height of this peak depends on the cleanliness of the superconductor; disorder scattering can smear the CdGM peak, leaving a LDOS that is flat over energy scales few times Δ in dirty samples[7,8]. Consequently, far from the VC, the LDOS exhibits a superconducting gap and coherence peak (partially broadened due to the orbital current), whereas inside the core the LDOS is either flat or exhibits a small peak at zero bias due to the formation of CdGM bound states in very clean samples. While the vast majority of superconductors follow the above description, a handful of superconductors, namely, High-$T_c$ cuprates[9,10] and strongly disordered NbN thin films[11,12] exhibit a radically contrasting behavior. In these systems inside the VC the LDOS is neither flat nor shows a zero energy peak but rather displays a LDOS suppression over energy scale comparable to the superconducting energy gap. While there is no consensus on the origin of this behavior, in high $T_c$ cuprates it has been suggested that it is alternatively associated with preformed Cooper pairs[9] or to the presence of a competing order[10]. In strongly disordered NbN thin films it has been recently suggested[13] that it is associated with the formation of Josephson vortices due to the formation of superconducting puddles induced by strong disorder.

In this paper, we report spectroscopic measurements in the vortex state of a weakly pinned thin film of the amorphous superconductor $Mo_{70}Ge_{30}$ (*a*-MoGe) using a low temperature scanning tunneling microscope. We observe that a soft gap in the LDOS continues to exist in the VC right up to the center of the vortex. Interestingly, *a*-MoGe is a conventional s-wave Type II superconductor with no known competing order. Furthermore, the superconducting state for the 20 nm thick film used in this study is very homogeneous as confirmed from STS spectroscopic maps in zero field, forcing us to look for explanations that have not been considered so far. We show that the magnetic field dependence of the soft gap can be explained if we assume that the vortices are not static but spatially fluctuate rapidly about their mean position. The spatial fluctuation at our lowest 450 mK is consistent with quantum zero-point fluctuation of vortices[14,15] which has been theoretically predicted in 2-dimensional (2D) superconductors.

## II. Sample and Experimental details

The sample used in this study consist of 20 nm thick *a*-MoGe film, similar to the ones used in refs.16,17,18. The superconducting transition temperature is, $T_c$ ~ 7.2 K. Earlier measurements showed that the very weakly pinned 2D vortex lattice undergoes two structural phase transitions with magnetic field, from a vortex solid to a hexatic vortex fluid (HVF) and from a HVF to an isotropic vortex liquid (IVL) [16]. As shown before the films have extremely weak pinning as evidenced from a very low depinning frequency and the absence of any difference between the field cooled and zero field cooled states.

STS measurements were performed in a home-built scanning tunneling microscope (STM) down to 450 mK and fitted with 90 kOe superconducting solenoid[19]. For STS measurements, post deposition, the film was transferred in the scanning tunneling microscope using an ultrahigh vacuum suitcase without exposure to air. Measurements were performed using a normal metal Pt-Ir tip. The differential tunneling conductance, *G(V)* was measured by superposing a small a.c. voltage, $V_{ac}$ (150 μV, 2.11 kHz) on the d.c. bias voltage $V_{dc}$ and measuring the a.c. current $I_{ac}$ using a lock-in amplifier, so that, $G(V = V_{dc}) = dI/dV|_{V=V_{dc}} \approx I_{ac}/V_{ac}$. To image the VL, *G(V)* maps were recorded with the bias voltage close to the superconducting coherence peak, such that each vortex appears as a local minimum in the conductance. The full *G(V)-V* spectra were recorded by switching off the feedback at a given location, and recording *G(V)* while the bias voltage is swept from positive to negative bias. The time to acquire a complete spectrum is ~ 2 seconds. While acquiring full spectroscopic area maps each spectra was averaged over three sweeps at every point.

## III. Results

We first characterize the zero-field properties of the *a*-MoGe film using STS. Fig. 1(a) shows the average *G(V)* vs. *V* tunneling spectra, acquired on a 32 × 32 grid over 200 nm × 150 nm area at different temperatures. The spectra are fitted with the tunneling equation[1],

$$G(V) = \frac{d}{dV}[\frac{1}{R_N}\int_{-\infty}^{+\infty} N_S(E)N_N(E-eV)\,(f(E)-f(E-eV)\}dE, \quad (1)$$

where $N_S(E)$ and $N_N(E) \approx 1$ are the normalized density of states for the superconducting and normal metal respectively, $f(E)$ is the Fermi-Dirac distribution function and $R_N$ is the resistance of the tunnel junction for $V \gg \frac{\Delta}{e}$. $N_S(E)$ is given by the BCS expression $N_S(E) = Re\left\{\frac{|E-i\Gamma|}{\sqrt{|E-i\Gamma|^2-\Delta^2}}\right\}$ where Γ is an additional phenomenological parameter[20] which accounts for all non-thermal sources of broadening. Fig. 1(b) shows the temperature variation of Δ and Γ along with the sheet resistance, $R_s$. *Δ(T)* follows the expected BCS variation and goes to zero exactly at the temperature where resistance appears. To produce the conductance maps we normalize the conductance curve at each point with its value at high bias, i.e. $G_N(V) = \frac{G(V)}{G(6\,mV)}$ .In Fig. 1(c)-(d) we plot the normalized conductance maps at zero bias, $G_N(0)$, and at the bias voltage corresponding to the coherence peak ( $V^p$ ) averaged for positive and negative voltages, $G_N^p = \frac{(G_N(V^p)+G_N(-V^p))}{2}$. Both the color plots as well as the very small spread in the conductance values observed from the histogram of $G_N(0)$ and $G_N^p$ (Fig. 1(e)) show the homogeneous nature of the superconducting state.

We now turn our attention to the spectroscopy of the vortex cores (Fig. 2). The top panels of Fig. 2(a)-(d) show the vortex lattice image in different fields at 450 mK over a small area containing 6-8 vortices. The panels in the second row show the normalized conductance spectra, $G_N(V)$ vs. $V$ along a line passing through the center of a vortex core. The panels in the third row show representative spectra at the center of the VC and approximately midway between two vortices. Though the superconducting gap starts to fill while approaching the vortex center, a soft gap in the tunneling spectra continues to exist even at the center of the vortex such that $G_N(0) < 1$. The bottom panels show the variation of $G_N(0)$ along three lines passing through the vortex center as shown by the dashed lines in the top panel. Here again we observe that $G_N(0) < 1$ even at the center of the vortex. In Fig. 2(e) we plot $G_N(0)$ at the center of the vortex and at the midpoint between two vortices as a function of magnetic field. We observe that $G_N(0)$ at the center of the vortex shows an overall increasing trend with magnetic field except for two anomalies around 5 kOe and 70 kOe. However, the soft gap does not close fully up to 80 kOe.

## IV. Discussion

*IV.A. Origin of the soft gap inside VC:* As mentioned before, the explanations for the soft gap in the VC given for High-$T_c$ cuprates or disordered NbN are contingent upon the presence of competing order or the presence of strong inhomogeneity in the superconducting state, neither of which are present in our *a*-MoGe film. Furthermore, we do not observe any characteristic feature such as a pseudogap state above $T_c$ which might suggest that the physics of preformed pair might play any role here. Thus, irrespective of whether those explanations are correct in the context they were proposed, we need to consider radically different explanation for the soft gap in the VC observed here.

We propose that the soft gap at the center of the vortex arises from rapid spatial motion of the vortices, which vibrate about their mean position. In a solid, which can be visualized as a coupled mass-spring system, such vibrations of atoms arise from two origins: (i) Thermal excitations and (ii) quantum zero-point fluctuations. As the solid is cooled, the thermal vibration of atoms gradually decreases and at low enough temperature the quantum zero point fluctuation sets the limiting value of atomic vibration at $T = 0$. For a vortex lattice where vortices play the role of atoms, the existence of lattice vibrations has been theoretically postulated many years back[21], even though they have not been experimentally demonstrated. Nevertheless, if we assume that the vortices are spatially vibrating about their mean position the presence of a soft gap at the center of the vortex can be naturally explained. Since STS is a slow measurement and probes the average tunneling conductance at a given location over time scale of few milliseconds, such rapid fluctuation will get integrated out, thus blurring the boundary between the VC and the gapped superconducting region in between the cores. Therefore the effect of fluctuation is that the tunneling conductance close to the center of the vortex would have contribution both from the VC as well as regions outside it, thus exhibiting the soft gap.

*IV.B. Simulating the conductance maps for fluctuating vortices:* We can now explore if the above-mentioned physical picture is consistent with our STS data. For that we need to simulate the conductance map with fluctuating vortices. We start with the simulation of a vortex lattice without spatial fluctuations. The exact calculation of the LDOS in the presence of vortices is in-principle possible by solving the Usadel equations[22], but this computation is very difficult[23]. Here, instead we adopt a phenomenological approach. First, we simulate the regular VL in a superconductor where the vortices are static. Since in *a*-MoGe the electronic mean free path is very short ($l$ ~ 0.15 nm), we assume that at the vortex core of an isolated vortex the CdGM peak is heavily smeared by disorder scattering[7], such that, $G_N(V) = 1$. Far from the core we assume that the conductance is given by the usual BCS relation given by eqn. (1), $G_N(V) = G_N^{BCS}(T,V,\Delta,\Gamma)$. $G_N^{BCS}(T,V,\Delta,\Gamma)$ is obtained by fitting the experimental zero field tunneling conductance spectra at the same temperature, with Δ and Γ used as fitting parameters. To interpolate between the two, we use an empirical Gaussian weight factor, $f(\boldsymbol{r}) = \exp[-\frac{r^2}{2\sigma^2}]$ ($\sigma$ is of the order of $\xi$), such that,

$$G_N(V,\boldsymbol{r}) = f(\boldsymbol{r}) + [1 - f(\boldsymbol{r})]G_N^{BCS}(T,V,\Delta,\Gamma) \qquad (2)$$

For a VL, the resultant normalized conductance, $\tilde{G}_N(V,\boldsymbol{r})$, is assumed to be a linear superposition of the conductance from all vortices, i.e.

$$\tilde{G}_N(V,\boldsymbol{r}) = \sum_i G_N(V,\boldsymbol{r}-\boldsymbol{r_i}) / [\sum_i G_N(V=0,\boldsymbol{r}-\boldsymbol{r_i})]_{max} \qquad (3)$$

where $\boldsymbol{r}_i$ is the position of the $i$-th vortex and sum runs over all vortices; the normalizing factor ensures that the $\tilde{G}_N(V=0) = 1$ at the center of each vortex. The consistency of this procedure has been checked by simulating the VL on an $NbSe_2$ single crystal ( see Appendix A ). Figure 3(a) show the simulated conductance plots at 10 kOe for $V=0$ (top panel) and $V = V^p = 1.45$ mV (middle panel).

We now take up the situation where the vortices are randomly fluctuating about their mean positions. We assume that the tunneling conductance measured by STS at each point is a temporal average of the instantaneous conductance due to fluctuating vortices. Thus to simulate the situation when the vortices are randomly fluctuating about their mean positions, we first compute using eqn.(3) the instantaneous conductance map, $\tilde{G}_N(V,\boldsymbol{r},\Delta a)$, for 200 realizations of a distorted hexagonal lattice, where each lattice point is displaced by a random vector, $\delta\boldsymbol{r}_i$, satisfying the constraint, $|\delta\boldsymbol{r_i}| \leq \Delta a$. Defined in this way, $\Delta a$ is the amplitude of fluctuations. Each of these 200 realization correspond to an instantaneous profile of the conductance. The final simulated conductance profile, $G_N^S(V,\boldsymbol{r},\Delta a)$, is obtained by taking the average of all 200 realization, i.e.

$$G_N^S(V,\boldsymbol{r},\Delta a) = \frac{1}{200}\sum_{L=1}^{200} \tilde{G}_N^{(L)}(V,\boldsymbol{r},\Delta a) = \langle \tilde{G}_N(V,\boldsymbol{r}) \rangle_{\Delta a} \qquad (4)$$

where the index $L$ runs over each of the 200 realizations of the distorted lattice. This procedure neglects any retardation effect, which in this context is equivalent to working in the Born-Openheimer approximation and is justified as long as the vibrational frequency of the vortices is much smaller than the gap frequency, $\Omega_0 = \frac{2\Delta(0)}{\hbar}$ $(\sim 3.89 \times 10^{12}\ Hz)$ . We will show later that it is the case here.

In Fig. 3(a)-(d), we show the outcome of these simulations at 10 kOe. Fig. 3(a) is for the case of no fluctuation, i.e. $G_N^s(V, \boldsymbol{r}, \Delta a = 0) \equiv \tilde{G}_N(V, \boldsymbol{r})$, whereas Fig. 3(b)-(d) are for increasing magnitude of fluctuations. The top panels show the spatial variation of $G_N^s(V = 0, \boldsymbol{r}, \Delta a)$ at 10 kOe. The middle panels show the corresponding conductance maps at the coherence peak voltage $G_N^s(V^p, \boldsymbol{r}, \Delta a)$. In the lower panels we plot the variation of $G_N^s(0, \boldsymbol{r}, \Delta a)$ and $G_N^s(V^p, \boldsymbol{r}, \Delta a)$ along a line passing through the center of the vortex core. At the center of the core ($x$ = 0), the $G_N^s(0, \boldsymbol{r}, \Delta a)$ progressively decreases from 1 with increase in $\Delta a/a$; at the same time $G_N^s(V^p, \boldsymbol{r}, \Delta a)$ remains larger than 1 showing incomplete suppression of the coherence peak.

*IV.C. Fitting the experimental data and extracting the fluctuation amplitude:* To fit the experimental data we take $\Delta a/a$ and $\Gamma$ as fitting parameters and constrain $\xi$ within 25% of the value obtained from[24] $H_{c2}$ ($\xi \sim$5.2 nm). Fig. 4(a)-(b) show the line-cuts of $G_N(0,r)$ along with the best fit $G_N^s(0, \boldsymbol{r}, \Delta a)$ at 2 different fields at 450 mK. (Fits for all magnetic field values are given in Appendix B.) Above 10 kOe, we use a larger $\Gamma$ compared to its zero field value to account for the additional broadening from the orbital current around vortices. While we primarily analyze the variation of $G_N(0,r)$, we note that the same set of parameters reproduces the qualitative variation of $G_N(V^p, r)$, but the conductance value is overestimated by 15-20%; this is most likely due to the fact that the phenomenological $\Gamma$ parameter does not capture suppression of the coherence peak from orbital supercurrent accurately. In Fig. 5(b) we plot $\Delta a/a$ as a function of magnetic field extracted at 450 mK. While $\Delta a$ decreases with increasing field, $\Delta a/a$ shows an overall increasing trend. There are two anomalies observed close between 5-7 kOe and around 70 kOe where we observe a sudden decrease. We will comment on this in the next subsection.

*IV.C. Theoretical analysis of vortex fluctuation:* To quantitatively understand the magnetic field variation of $\Delta a/a$, we now theoretically compute this quantity within harmonic approximation. The lattice vibration of the 2D vortex lattice is governed by two elastic moduli[25]: compression, $C_{11}$, and shear $C_{66}$. However $C_{66} << C_{11}$, and therefore for small deformation the elastic energy of the VL is mainly controlled by $C_{66}$. In the language of lattice vibrations this means that transverse lattice vibration modes are easier to excite than longitudinal ones. Since the vortex motion is overdamped we assume that each vortex oscillates individually. Therefore instead of focusing on the full dispersion relation we use a simple Einstein model of vortex lattice phonon as has been done in ref. 14. It is important to note that even though this analysis is formulated assuming a vortex solid, it will remain valid even in a vortex fluid as long as there are short range hexagonal coordination and the diffusive motion of the vortices is much slower compared to the Einstein frequency.

To start with let us assume that vortices in a row are displaced by a distance δ (<< *a*), from its equilibrium position as shown in Fig. 5(a). The shear strain in the lattice,

$$\tan(\theta) \approx \theta = \frac{\delta}{a} \tag{5}$$

The corresponding elastic energy per unit volume,

$$\varepsilon = \frac{1}{2}\, C_{66} \left(\frac{\delta}{a}\right)^2 \tag{6}$$

So, the total elastic energy per vortex, $E = \varepsilon A d = \frac{1}{2}\, C_{66} \left(\frac{\delta}{a}\right)^2 A d$, where A is area of rhombus unit cell and $d$ is the thickness of film. In the London limit, the shear modulus of an isotropic superconductor for small magnetic field is given by[25],

$$C_{66} = \frac{\emptyset_0 H}{16\pi\mu_0\lambda^2} \tag{7}$$

Using this expression, the elastic energy is,

$$E^{el} = \frac{\emptyset_0 d}{32\pi\mu_0\lambda^2}(HA)\left(\frac{\delta}{a}\right)^2 \quad (8)$$

Since we have one vortex per unit cell, $HA = \Phi_0$,

$$E^{el} = \frac{{\emptyset_0}^2 d}{32\pi\mu_0\lambda^2}\left(\frac{\delta}{a}\right)^2 = \frac{1}{2}\left(\frac{{\emptyset_0}^2 d}{16\pi\mu_0\lambda^2 a^2}\right)\delta^2. \quad (9)$$

Therefore, motion of vortices about its equilibrium position, is analogous to a harmonic oscillator with Einstein frequency,

$$\omega = \sqrt{\frac{{\emptyset_0}^2 d}{16\pi\mu_0\lambda^2 a^2 m_v}} = \left(\frac{K}{m_v}\right)^{1/2}\frac{1}{a} = \omega_0\left(\frac{H}{H_{c2}}\right)^{1/2}, \quad (10)$$

where, $m_v$ is vortex mass and $K = \frac{{\emptyset_0}^2 d}{16\pi\mu_0\lambda^2}$ and $\omega_0 = \frac{1}{1.075}\left(\frac{KH_{c2}}{m_v\Phi_0}\right)^{1/2}$; we have used $a \approx 1.075\left(\frac{\Phi_0}{H}\right)^{\frac{1}{2}}$ to get the last form.

The total energy for each vortex oscillating with amplitude $\Delta a$ is given by the energy of the simple Harmonic oscillator, $m_v\omega^2(\Delta a)^2$. To estimate the quantum zero point motion we equate this quantity to the zero point energy, $\hbar\omega$, which gives,

$$\frac{\Delta a}{a} = \left(\frac{\hbar^{\frac{1}{2}}}{(Km_v)^{1/4}}\right)\left(\frac{1}{a}\right)^{1/2} \propto H^{\frac{1}{4}}. \quad (11)$$

This expression is very similar to a more detailed calculation in ref. 15 and differs only by factor of 0.6 ( see Appendix C ). On the other hand if the motion is thermal in origin, $m_v\omega^2(\Delta a)^2 \sim k_B T$. In this case, $\frac{\Delta a}{a} \sim \left(\frac{k_B T}{K}\right)^{1/2}$, which is independent of magnetic field.

In Fig. 5 (b) we show the $\frac{\Delta a}{a}$ vs. $H$ variation at 450 mK and 2K. We see that the $\frac{\Delta a}{a}$ vs. $H$ variation at 450 mK (black points) is very well captured with $H^{1/4}$ dependence (red line), consistent with quantum zero point motion. From the coefficient of the fit and using the experimental value of $\lambda \approx 534\, nm$ we obtain $m_v \sim 36m_e$ (where $m_e$ is the electron mass). Though different estimates of $m_v$ considerably vary[26,27,28,29,30,31] we can compare this value with the most widely used estimate[26], $m_v = \frac{2}{\pi^3} m^* k_F d$, where $m^*$ is the effective mass and $k_F$ is the Fermi wave vector. Assuming $m^* = m_e$, and using the free electron expression, $k_F = (3\pi^2 n)^{1/3}$, where $n$ = 5.2×10$^{29}$ el/m$^3$ is the carrier density determined from Hall effect, we obtain $m_v \approx 32m_e$ which is in very good agreement with the value obtained from our experiments. On the other hand the corresponding value of $\omega_0 = 8.65 \times 10^{11}\ Hz < \Omega_0$ justifying the Born-Oppenheimer approximation used section *IV.B*. Coming to the cusp-like anomalies observed at between 5-7 kOe and 70 kOe we note that these anomalies appear very close to the vortex solid to HVF, and HVF to IVL boundaries respectively. These anomalies most likely arise from the anharmonicity of the confining potential close to the phase boundaries, though a detailed understanding will require further investigations. It is interesting to note that the first anomaly appears when $\frac{\Delta a}{a} \sim 0.14$, which is in the ballpark value expected from the Lindemann criterion for melting.

Coming to the $\frac{\Delta a}{a}$ vs. $H$ data at 2K, we note that the value is very close to the 450 mK value, for magnetic fields above 40 kOe but becomes larger than the 450 mK value at lower fields. To understand this behavior we need to account for the role of thermal fluctuation. The quantum to thermal crossover is expected to happen when the thermal excitation energy exceeds the oscillator level spacing, i.e. $k_B T \gtrsim \hbar\omega = \hbar\omega_0 \left(\frac{H}{H_{c2}}\right)^{1/2}$. Thus at a given temperature thermal fluctuations will dominate when $H \lesssim H_{c2} \left(\frac{k_B T}{\hbar\omega_0}\right)^2 \equiv H_{cross}$. At 450 mK this gives $H_{cross} = 0.54\ kOe$ ($H_{c2}$ ~ 126 kOe); since the lower limit of our data is 1 kOe thermal fluctuations can be entirely neglected. To estimate $H_{cross}$ at 2 K we can use the same value of $\omega_0$ since the λ does not change significantly in this temperature range. Using $H_{c2}$ ~ 108 kOe at 2K we obtain $H_{cross} \sim 9.5\ kOe$. Consequently, much above this field $\frac{\Delta a}{a}$ is entirely dominated by quantum fluctuations for both 450 mK and 2 K. However, as the field is reduced $\frac{\Delta a}{a}$ at 450 mK continues to be dominated by quantum fluctuations whereas at 2 K the additional magnetic field independent contribution of thermal fluctuations becomes significant. This is consistent with our observation even though a quantitative fit of the data under the combined effect of quantum zero point motion and thermal fluctuations in beyond the scope of the present analysis.

## V. Conclusion

We have shown evidence of quantum zero-point fluctuation of vortices at low temperatures in a 2D vortex lattice in a weakly pinned $a$-MoGe thin film. The spatial fluctuation of vortices leaves a very clear signature on the spectral property of the vortex in the form of a soft gap at the center of the vortex. Given the characteristic frequency of fluctuation from our analysis ( $\omega_0 \approx 0.865$ THz ) it might be worthwhile to look for direct signatures of vortex fluctuation from THz measurements. It would also be interesting to investigate the impact of zero-point fluctuations on the state of the 2D vortex lattice at low temperatures. In liquid $^4$He the zero point fluctuation prevents the liquid from solidifying and thus producing a quantum fluid at $T = 0$. Whether zero point fluctuations can produce a quantum vortex fluid at very low temperatures is at present unclear. Even though existing studies indicate that the hexatic and isotropic vortex fluid continue to exist over a large magnetic field range down to 450 mK, it would be interesting to explore this more carefully through transport and STS imaging down to lower temperatures.

## Appendix A

### Simulating the conductance map for a vortex lattice

To cross-check the validity of the phenomenological model used to simulate the conductance maps for a vortex lattice, we apply it on a conventional system, namely the vortex state in clean 2H-$NbSe_2$ (single crystal), where we do not have any evidence that the vortices fluctuate about their mean positions. Fig. 6(a) shows the VL image on a $NbSe_2$ single crystal at 5 kOe, 450 mK, keeping the bias voltage at 1.2 meV, close to the coherence peak. Fig. 6(b) shows the $G_N(V)$ vs $V$ spectra along a line passing through the center of the vortex. Here, the core of vortex shows a zero bias conductance peak ($G_N(V=0) > 1$) resulting from the bound state of normal electrons inside the normal core, known as Caroli-de Gennes-Matricon bound state[3]. On the other hand, spectra obtained at superconducting regions away from the vortex core has regular BCS characteristics, partially broadened by the circulating current around the vortex core (*inset* Fig. 6(c)). The experimental variation of $G_N(0)$ along three lines passing through the center of the vortex as well as their average is shown in Fig. 6 (c).

The VL is simulated as follows. We assume that far away from the vortex core $G_N(V)$ will be similar to the conductance spectra in zero field. This is obtained by fitting the zero field experimental tunneling conductance with BCS theory using superconducting energy gap, $\Delta$ and phenomenological broadening parameter, $\Gamma$ as the fitting parameters; we call this resultant best fit spectrum as $G_N^{BCS}(V)$. For the spectra at the center at the vortex center, $G_N^{center}(V)$ we used the average of the experimental spectra obtained at the center of the several vortices. These two spectra are shown in Fig. 6 (d). To interpolate between these two, we use an empirical Gaussian weight factor, $f(\boldsymbol{r}) = \exp\left(-\frac{\boldsymbol{r}^2}{2\sigma^2}\right)$, such that $G_N(V,\boldsymbol{r}) = f(\boldsymbol{r})G_N^{center}(V) + [1 - f(\boldsymbol{r})]G_N^{BCS}(V)$, where $\boldsymbol{r}$ is the position with respect to the center of the vortex. We choose $\sigma \approx \xi$ and where $\xi \sim 8.9\ nm$ (corresponding to the measured $H_{c2} \sim 42$ kOe) is the Ginzburg-Landau coherence length of clean $NbSe_2$. Using this we construct the VL by linear superposition of conductance values from all vortices and henceforth obtain resultant conductance map given by, $\tilde{G}_N(V,\boldsymbol{r}) = 1.6 * \sum_i G_N(V,\boldsymbol{r}-\boldsymbol{r_i}) / [\sum_i G_N(V=0,\boldsymbol{r}-\boldsymbol{r_i})]_{max}$ where position of *i*-th vortex is $\boldsymbol{r}_i$. This construct ensures that at the center of the vortex the simulated conductance matches with the experimental zero bias conductance at the vortex center. In Fig. 6(e) we compare the experimental variation of $G_N(0)$ for a line passing through the vortex center along with the corresponding value obtained from our simulation. The *inset* shows the corresponding data for $G_N(V^p = 1.2$ meV$)$ which is close to the superconducting coherence peaks. The good agreement in both cases shows the validity of our phenomenological approach to simulate the conductance map.

One difference between the above simulation and the one in the case in $a$-MoGe is that in that case we take $G_N^{center}(V) = 1$. The reason is that zero bias conductance peak at the center of the vortex core is very sensitive to disorder and very small amount of scattering (such as small amount of Co doping in $NbSe_2$) destroys the peak giving a flat spectrum[32]. Therefore it is unlikely to be present even in the absence of fluctuations in $a$-MoGe where the electronic mean free path is very small due to the amorphous nature of the sample.

## Appendix B

**Fit of the *$G_N(0)$* profile passing through the vortex core.**

In Fig. 4 we showed two representative fits of our model with $G_N(0)$ profiles passing through the vortex core. In Fig. 7 (a)-(h) we show the fits at several other magnetic fields (at 450 mK), along with the best fit parameters. Fig. 7 (i) shows the value of Γ used to fit the profiles. Above 10 kOe we need to gradually increase the value of Γ to take into account the broadening arising from orbital current around the vortex core.

## Appendix C

In this section, we will provide a detailed calculation to show that the expression of $\Delta a/a$, given by eqn. (11) is similar to the expression derived in ref. 15. We start with eqn. 33 of reference 15, which in the harmonic limit reduces to,

$$m_v = \frac{0.03627 a_v^2 \hbar^2}{\rho_s u_{rms}^4}, \qquad (12)$$

where $a_v \equiv a$ is the separation between nearest neighbor vortices, $u_{rms}^2$ is the mean square displacement. $\rho_s$ is the superfluid stiffness and it is denoted as,

$$\rho_s = \frac{\hbar^2 c^2 d}{16\pi e^2 \lambda^2} = \frac{\hbar^2 d}{4\mu_0 e^2 \lambda^2} = \frac{4}{\pi} K \qquad (13)$$

From eqn. (12) we get,

$$u_{rms}^4 = \frac{0.03627 a_v^2 \hbar^2}{\rho_s m_v} \Rightarrow \frac{u_{rms}}{a_v} = \left(\frac{0.03627\hbar^2}{\rho_s m_v}\right)^{1/4} \frac{1}{\sqrt{a_v}} \qquad (14)$$

Now using eqn. (13) and noting that the amplitude of oscillation ( $\Delta a = \sqrt{2} u_{rms}$). we can write eqn. (14) as,

$$\frac{\Delta a}{a} = \sqrt{2}\left(\frac{0.03627\hbar^2}{\frac{4}{\pi} K m_v}\right)^{1/4} \left(\frac{1}{a}\right)^{1/2} = 0.6\frac{\hbar^{1/2}}{(K m_v)^{1/4}}\left(\frac{1}{a}\right)^{1/2} \qquad (15)$$

Eqn. (15) identical to equation (11) except for the factor of 0.6.

*Acknowledgements:* The work was supported by Department of Atomic Energy, Govt. of India (Grant No. 12-R&D-TFR-5.10-0100) and U.S. National Science Foundation (Grant No. DMR-2002850).

SD and IR performed the STS measurements and analyzed the data. JJ optimized deposition conditions and synthesized the samples. The theoretical analysis were carried out by SD, SS and PR. PR conceived the problem, supervised the experiments and wrote the paper. All authors read the manuscript and commented on the paper.

---

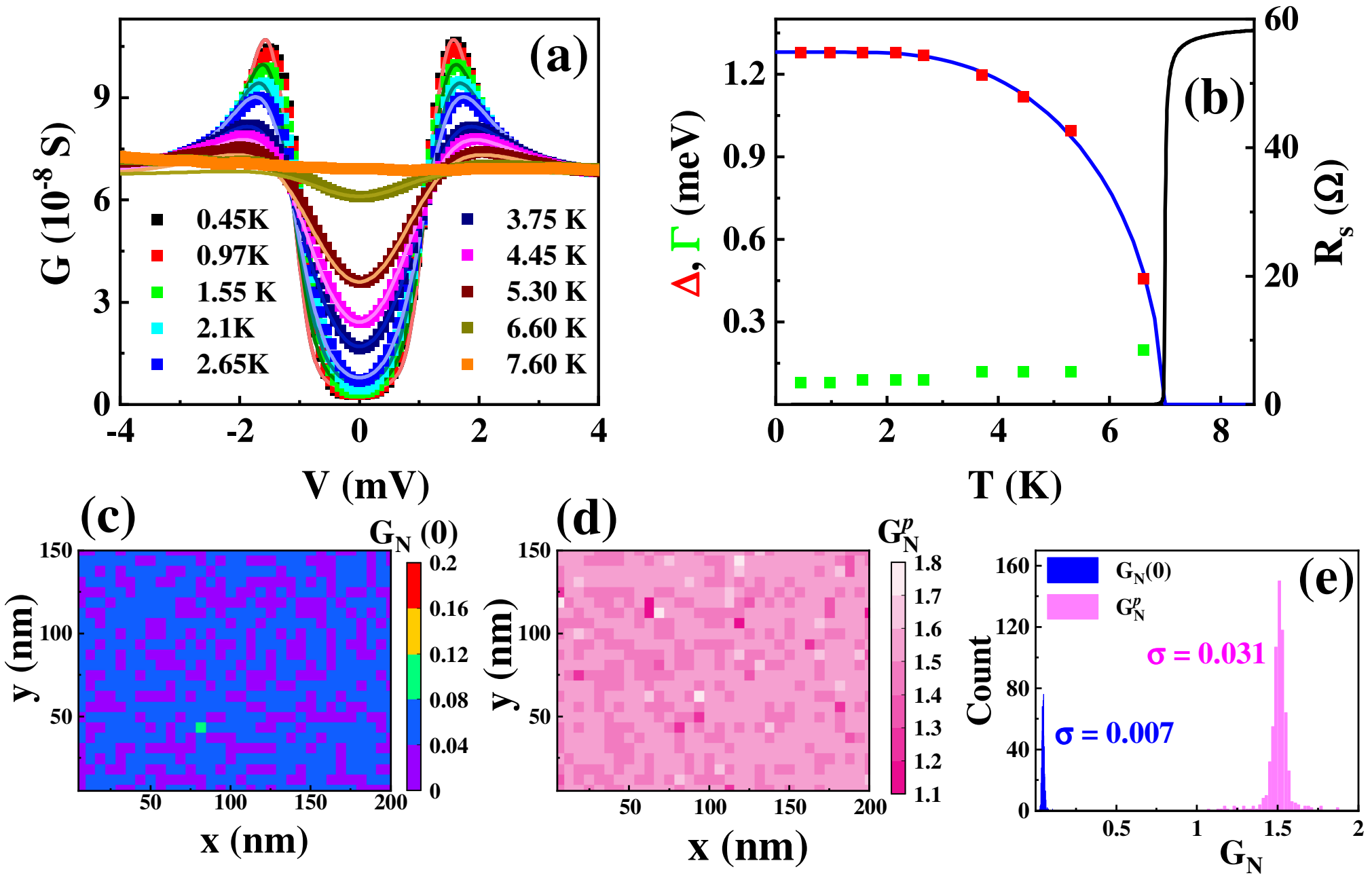

**Figure 1|** (a) $G(V)$-$V$ tunneling conductance spectra in zero field at different temperatures; the solid lines are the fit with eqn. (1). (b) Temperature variation of Δ and Γ obtained from the fits; the temperature variation of $R_s$ is shown in the same panel; the solid blue line is the expected BCS variation of Δ. Normalised conductance maps at (c) Zero bias and (d) bias voltage corresponding to the coherence peak, $V^p = 1.45\ mV$. (e) Histogram of conductance values, $G_N(0)$ and $G_N^p$, corresponding to panels (c) and (d); the standard deviations (σ) of $G_N(0)$ and $G_N^p$ are shown next to the histograms.

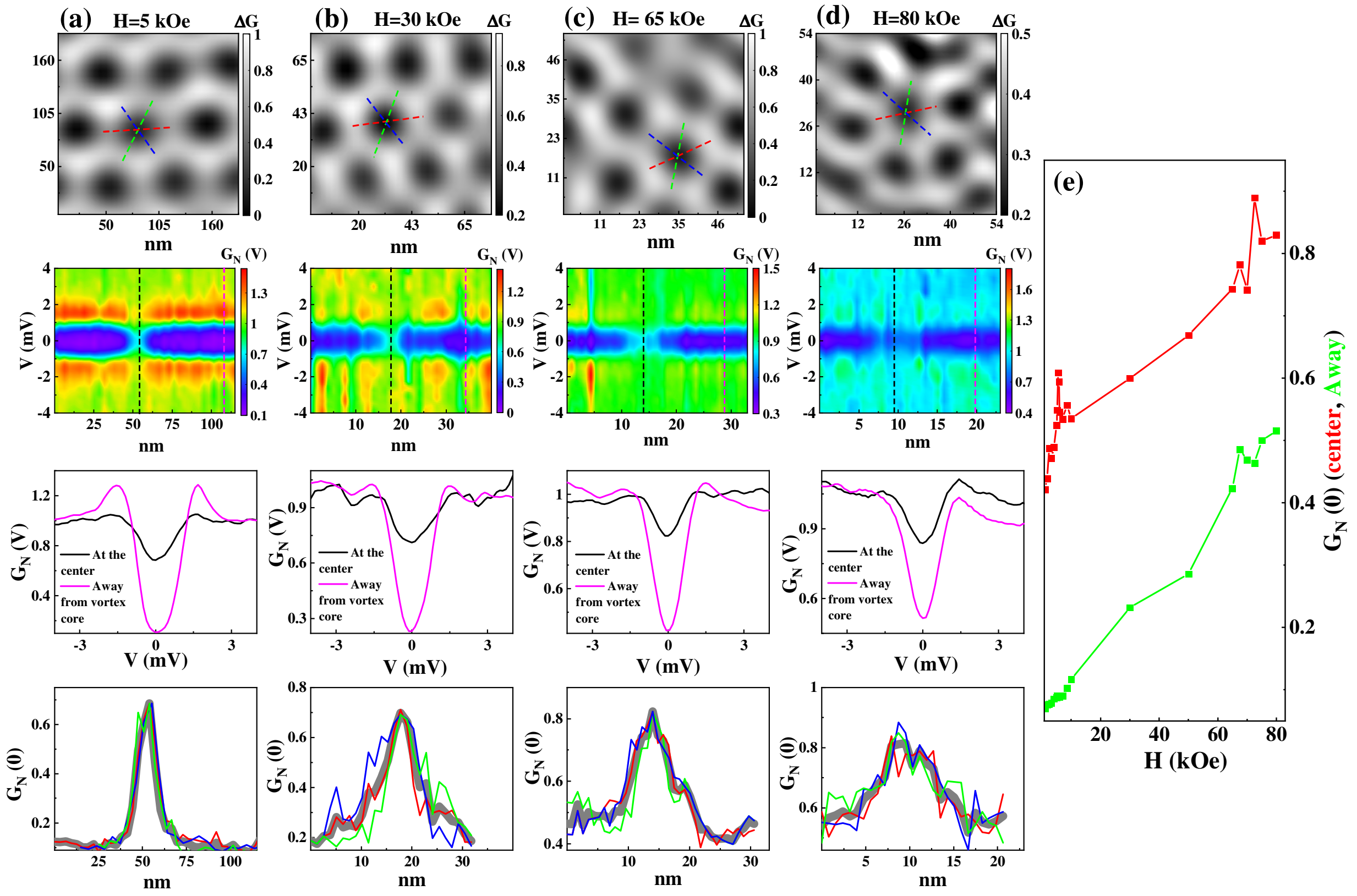

**Figure 2**| (a)-(d) The top panels show the image of the vortex lattice over a small area at 450 mK in a field of 5 kOe, 30 kOe, 65 kOe and 80 kOe respectively; here the local variation is conductance, ΔG, is in arbitrary units. The second row show the intensity plot of $G_N(V)\ vs. V$ along a line passing through the center of a vortex shown by the red dashed lines in the top panels. The third row show the spectra at the center of a vortex (black) and at the midpoint between two vortices (purple) corresponding to the vertical dashed lines of the same color in the second row. The bottom panels show spatial variation of $G_N(0)$ along the three dashed lines passing through the center of a vortex in the top panel respectively along with their average (thick grey line); (e) Variation of $G_N(0)$ with magnetic field at the center of vortices and far from the center.

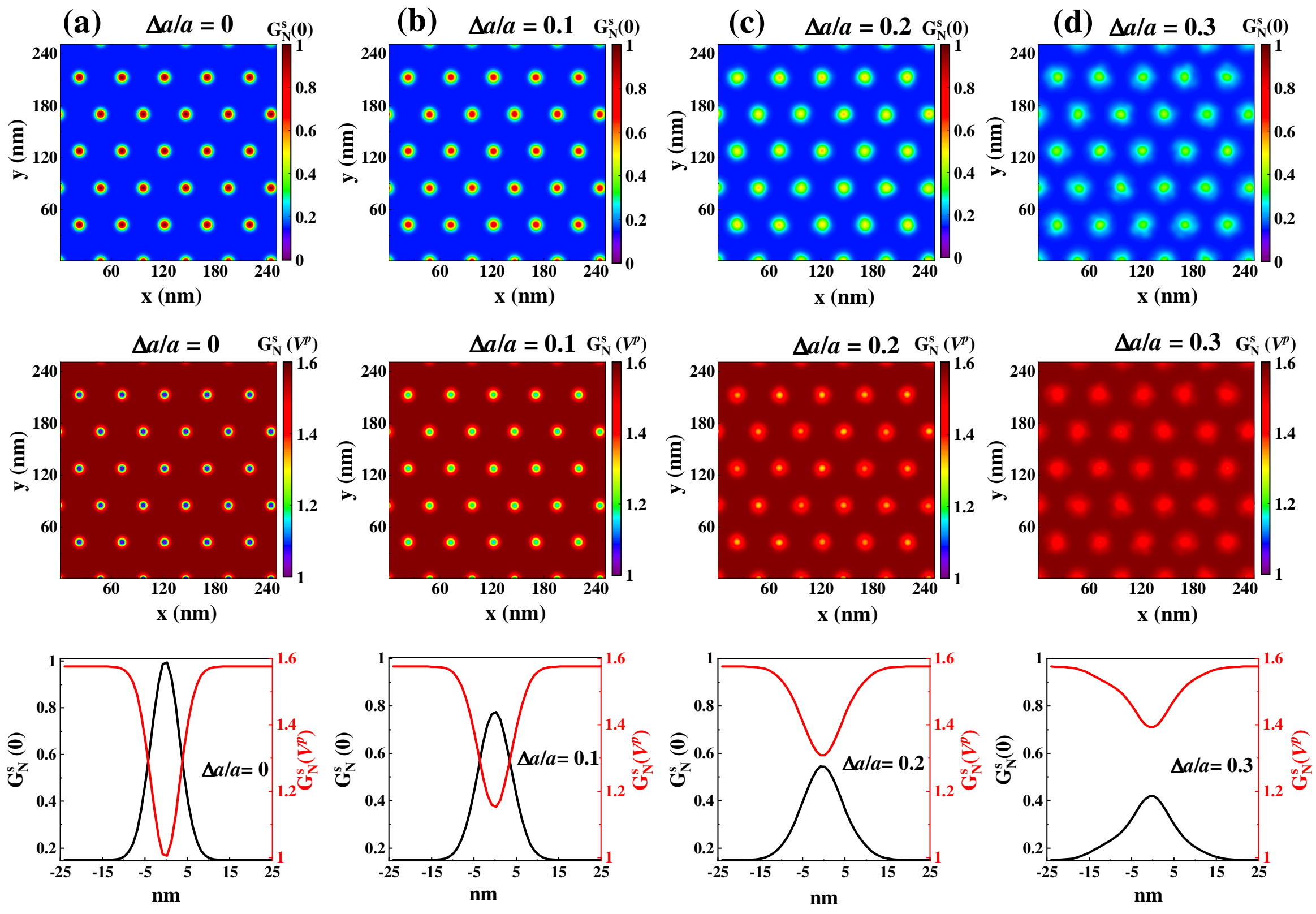

**Figure 3**| Representative simulation of the normalized conductance at 10 kOe, 450 mK for different fluctuation amplitudes: (a) $\frac{\Delta a}{a} = 0$, (b) $\frac{\Delta a}{a} = 0.1$, (c) $\frac{\Delta a}{a} = 0.2$, (d) $\frac{\Delta a}{a} = 0.3$. The top panels show the normalized conductance maps for $V = 0$ ($G_N^s(0) \equiv G_N^s(0, \boldsymbol{r}, \Delta a)$). The middle panels show the normalized conductance maps for $V = V^p = 1.45$ mV ($G_N^s(V^p) \equiv G_N^s(V^p, \boldsymbol{r}, \Delta a)$). The bottom panels show the variation of $G_N^s(0)$ and $G_N^s(V^p)$ along a line passing through the center of the vortex.

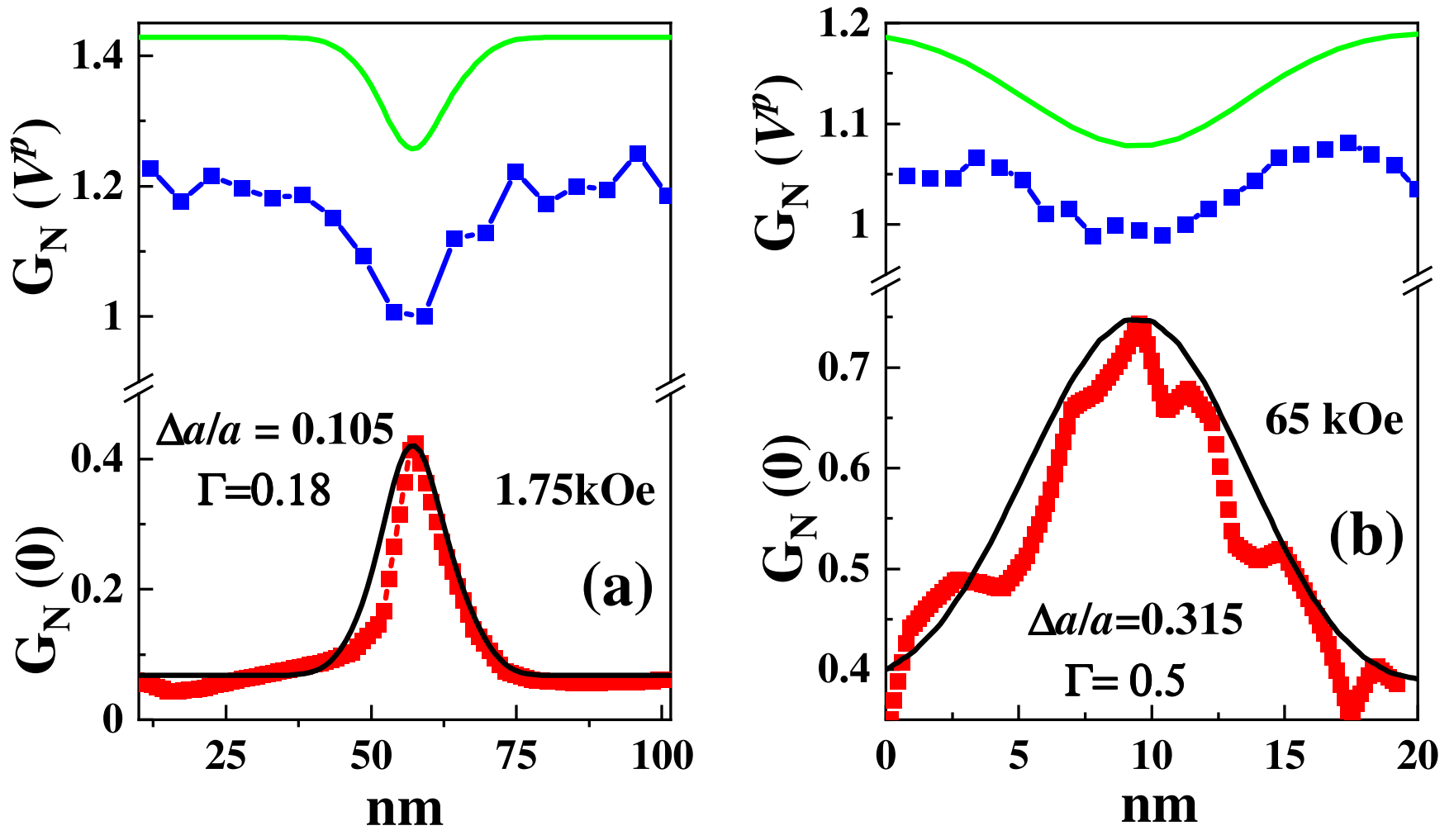

**Figure 4|** (a)-(b) Spatial variation of $G_N(0)$ (Red points) and $G_N(V^p)$ (Blue points) along with corresponding theoretical fits (black and green lines) for 2 different magnetic fields; the spatial variation is obtained by averaging over 3 lines passing through the center of a vortex. The magnetic field value and the best fit values of $\frac{\Delta a}{a}$ and Γ corresponding to each panel is shown in the legend.

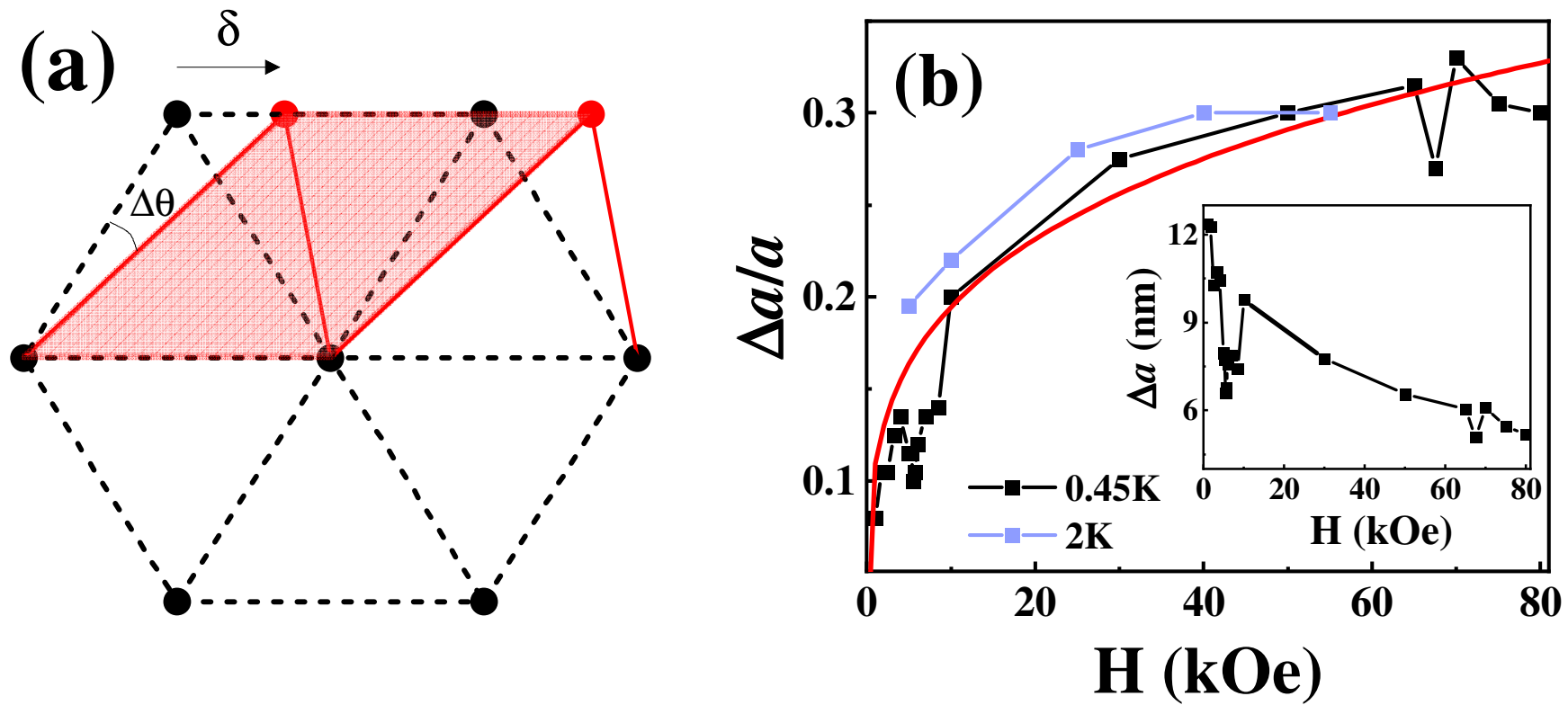

**Figure 5|** (a) The black points show one undistorted hexagonal unit cell; the red points show the positions of the top two lattice points after applying a shear deformation (of exaggerated magnitude). The shaded red region shows the distorted primitive unit cell. (b) $\frac{\Delta a}{a}$ as a function of magnetic field at 450 mK (connected black points) and 2 K (connected blue points); the red line is a fit of the 450 mK data to $\frac{\Delta a}{a} \propto H^{\frac{1}{4}}$. The inset shows $\Delta a$ as a function of field at 450 mK.

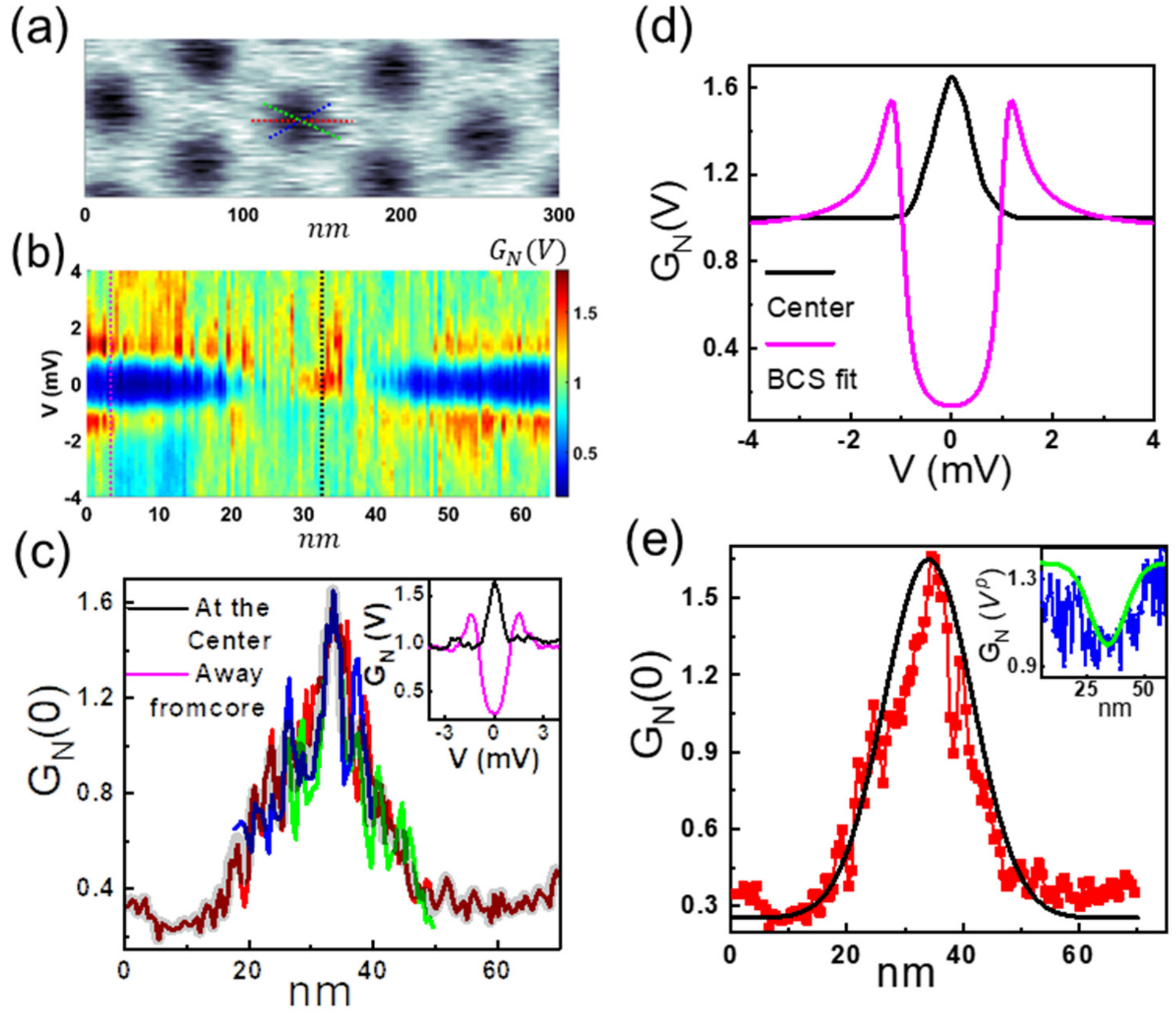

**Figure 6|** (a) Vortex image on $NbSe_2$, observed by recording $\Delta G(V = 1.2\ meV)$ across the area, showing three lines passing through a vortex center. (b) $G_N(V)$ vs. $V$ spectra along the red dotted line in (a) with black dotted line denoting the center of the vortex. (c) Variation of $G_N(0)$ along the dotted lines passing through the vortex cores in (a); *inset* Two spectra at the black and pink dotted lines in (b) respectively. The transparent grey is the average of the three. (d) The two spectra at the center of the vortex core and far away from the vortex core taken for the purpose of our simulation. (e) Simulated $G_N(0)$ (black line) along with the average $G_N(0)$ (red squares) for three lines passing through the center of the vortex obtained from experimental data. (*inset*) Simulated $\tilde{G}_N(V = 1.2\ meV)$ (green line) along with the average $G_N(V = 1.2\ meV)$ (blue line) obtained from experimental data.

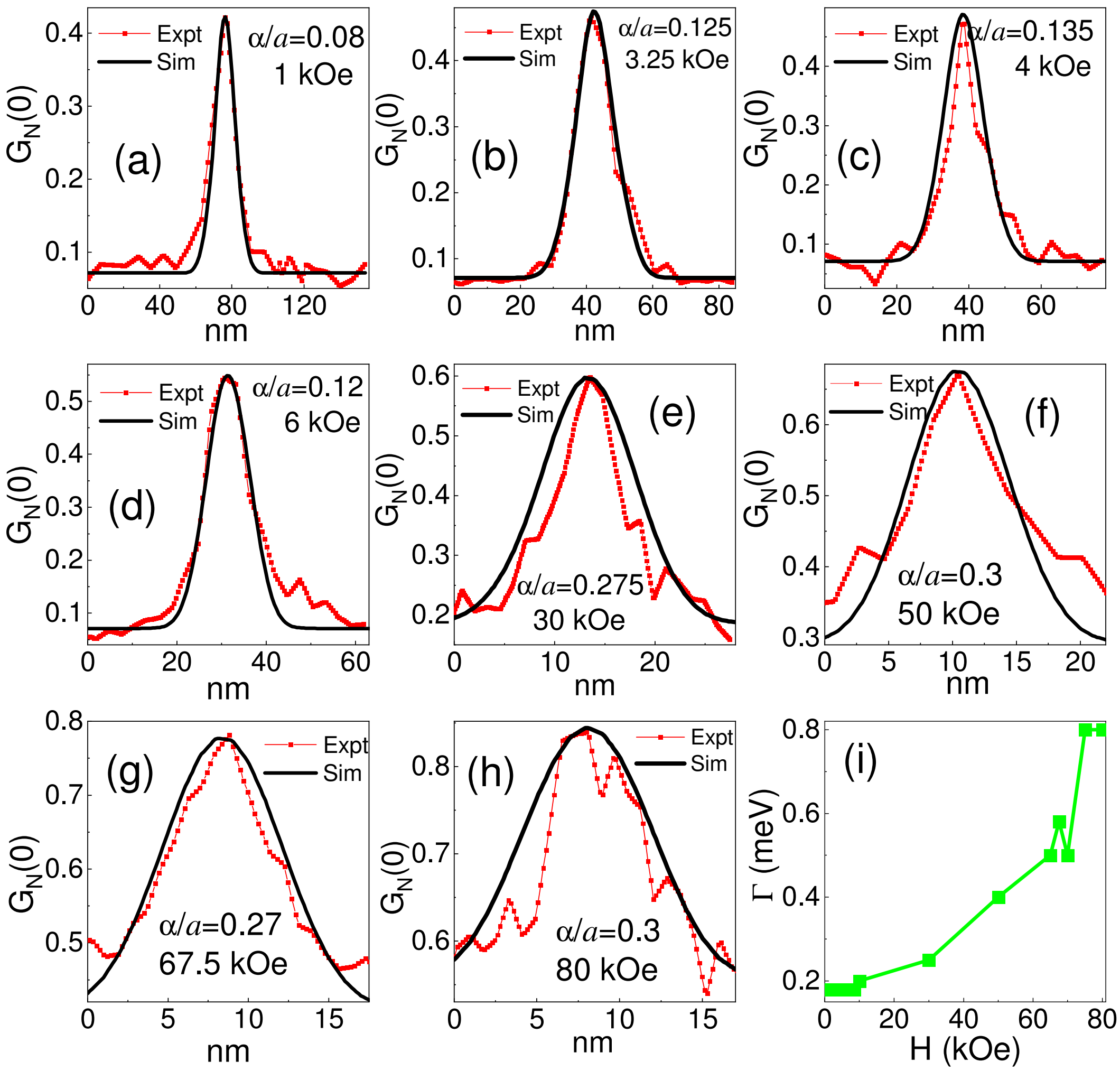

**Figure 7|** (a)-(g) The red connected points are averaged line-cuts of $G_N(0,r)$ along vortex cores in different magnetic fields. The black lines are line-cuts of simulated $G_N^s(0,r)$ at the corresponding field. (i) Magnetic field variation of the broadening parameter Γ used in the fits.